\documentstyle[12pt]{article}
\title{\baselineskip=9mm
Path integral approach to no-Coriolis approximation
in heavy-ion collisions}
\author{K. Hagino,$^1$ N. Takigawa,$^1$, A.B. Balantekin$^2$, and
J.R. Bennett$^3$
\\ \\
\medskip
{\it $^1$ Department of Physics,
Tohoku University, 980--77 Sendai, Japan}\\
{\it $^2$ Physics Department, University of Wisconsin, }\\
{\it Madison, Wisconsin~53706, USA }\\
{\it $^3$ Department of Physics and Astronomy,}\\
{\it University of North Carolina at Chapel Hill,
 Chapel Hill, NC 27599--3255}
}
\date{}

\begin{document}
\baselineskip=9mm
\maketitle

\begin{center}
{\bf Abstract}
\end{center}

We use the two time influence functional method of the
path integral approach in order to reduce the dimension of
the coupled-channels equations for heavy-ion reactions
based on the no-Coriolis approximation.
Our method is superior to other methods in that it
easily enables us to study the cases where
the initial spin of the colliding particle is not zero.
It can also be easily applied to the cases
where the internal degrees of freedom are not necessarily
collective
coordinates.
We also clarify the underlying assumptions in
our approach.

\medskip

\noindent
PACS number(s):
25.70.Jj, 74.50.+r, 21.10.Re

\medskip

\newpage

It is by now well established that heavy-ion fusion reactions
at energies below the Coulomb barrier are not such simple processes
that can be described in terms of penetration through a
one dimensional potential barrier, but rather complicated reactions
where internal degrees of freedom of colliding nuclei
play an essential role\cite{B88}.
Therefore they are
typical examples of macroscopic quantum tunneling,
which has been a very popular subject in the past decade
in many subfields of physics and chemistry[2--4].
%\cite{CL81,HTB90,JJAP93}.
One of the major interests in macroscopic
quantum tunneling is to assess the effects of the environment on the
tunneling rate of a macroscopic  degree of freedom.
A standard way to tackle this problem in nuclear physics
is to numerically solve the associated coupled-channels equations.
However, the full coupled-channels calculations quickly become
very intricate if many physical channels are included. This makes
an intuitive understanding of the numerical results quite hard.
For this reason, an approximation named the no-Coriolis
approximation, which is sometimes called the rotating frame approximation,
has been introduced[5--11].
%\cite{TI86,NBT86,T87,ELP87,TAB92,AT93,GCJ86}.
It is a sudden tunneling approximation\cite{HTBB95} concerning
the centrifugal energy and it
greatly reduces the number of coupled channels to be solved.
The no-Coriolis approximation was first introduced
in the field of chemistry under the name centrifugal sudden
approximation[13--15].
%\cite{MS84,MK74,MTN94}.
Recently, it has also been applied to the problem of
electron-molecule scattering\cite{ALS93}.

The no-Coriolis approximation has been derived by several different methods.
The authors in ref.[5--8]
%\cite{TI86,NBT86,T87,ELP87}
used properties of Racah coefficients, and ref. \cite{TAB92,AT93}
used the Green's function method. Symmetry considerations using tidal
spin have been used in ref. \cite{GCJ86}.
The aim of this paper is to present a new derivation of the
no-Coriolis approximation using the path integral method\cite{BT85}.
This approach had already been used in ref.\cite{BBT91},
but the angular momentum coupling was not treated explicitly.
A salient point of our derivation is that it
enables us to
easily extend to the cases where
the initial spin of the colliding particles is not zero, and
where there is a spin-orbit force in the scattering process.
It can also be easily applied to the cases where the
internal degrees of freedom are not collective coordinates, but
the coordinates of the constituent particles of the colliding system.
It also clarifies the underlying assumption of the approximation.

We consider the collision between two nuclei in the presence of
the coupling between the translational motion, i.e. the relative motion
between the centers of mass of the colliding nuclei,
${\mbox{\boldmath $R$}}=(R,\Omega)$ and a
nuclear intrinsic motion $\xi$.
We assume the following Hamiltonian for this system
\begin{equation}
H({\mbox{\boldmath $R$}},\xi)=-\frac{\hbar^2}{2\mu}\nabla^2+U(R)+H_0(\xi)
+V({\mbox{\boldmath $R$}},\xi)
\end{equation}
where $\mu$ is the reduced mass.
$U(R), H_0(\xi)$ and $V({\mbox{\boldmath $R$}},\xi)$ are the bare potential
energy for the translational motion, the internal Hamiltonian and
the coupling Hamiltonian, respectively.
In general the internal degree of freedom $\xi$ has a finite spin.
We therefore expand the coupling Hamiltonian in multipoles
\begin{equation}
V({\mbox{\boldmath $R$}},\xi)
=\sum_{\lambda>0}f_{\lambda}(R)Y_{\lambda}
(\Omega)\cdot
T_{\lambda}(\xi).
\end{equation}
Here $Y_{\lambda}(\Omega)$ are the spherical harmonics and
$T_{\lambda}(\xi)$ are spherical tensors constructed from
the internal coordinate. The dot indicates a scalar
product.
The sum is taken over all values of $\lambda$, except for $\lambda=0$ which
is already included in $U(R)$.

For a fixed total angular momentum $J$ and its $z$ component $M$,
the expansion basis of the
coupled-channels equations are defined as
\begin{equation}
<\Omega\xi\vert(nLI)JM>
=\sum_{m_L,m_I}<Lm_LIm_I\vert JM>
Y_{Lm_L}(\Omega)\varphi_{nIm_I}(\xi),
\end{equation}
where $L$ and $I$ are the orbital and the internal
angular momenta, respectively.
$\varphi_{nIm_I}(\xi)$ are the wave functions of the internal motion
which obey
\begin{equation}
H_0(\xi)\varphi_{nIm_I}(\xi)=\epsilon_{nI}\varphi_{nIm_I}(\xi).
\end{equation}
If we expand the total wave function with this basis as
\begin{equation}
\Psi_{J}({\mbox{\boldmath $R$}},\xi)=
\sum_{n,L,I}\frac{u^{J}_{nLI}(R)}{R}
<\Omega\xi\vert(nLI)JM>,
\end{equation}
the coupled-channels equations for $u^{J}_{nLI}(R)$ read
\begin{equation}
\left[-\frac{\hbar^2}{2\mu}\frac{d^2}{dR^2}
+\frac{L(L+1)\hbar^2}{2\mu R^2}+U(R)
-E+\epsilon_{nI}\right]u^{J}_{nLI}(R)
+\sum_{n',L',I'}V^{J}_{nLI; n'L'I'}(R)u^{J}_{n'L'I'}=0
\label{B.7}
\end{equation}
with
$V^{J}_{nLI ; n'L'I'}(R)=<JM(nLI)\vert V({\mbox{\boldmath $R$}},\xi)
\vert (n'L'I')JM>$ . We have suppressed the index $M$ in
$V^{J}_{nLI ; n'L'I'}(R)$, since they are independent of that
quantum number.
These coupled-channels equations are solved with the  boundary conditions
\begin{equation}
u^{J}_{nLI}(R)\to \frac{1}{T^J_{nLI}}H_L^{(-)}(R)\delta_{n,n_i}
\delta_{L,L_i}\delta_{I,I_i}+\frac{R^J_{nLI}}{T^J_{nLI}}H_L^{(+)}(R)
{}~~~~(R\to\infty)
\end{equation}
where $H_L^{(+)}(R)$ and $H_L^{(-)}(R)$ are the outgoing and the
incoming Coulomb waves, respectively.
Once the coefficients in the asymptotic region $T^J_{nLI}$ are obtained,
the penetration probability through the
Coulomb potential barrier is given by
\begin{equation}
P^{J}_{L_iI_i}(E)=
\sum_{n,L,I}
\frac{k_{nI}}{k_{n_iI_i}}\vert T^J_{nLI} \vert ^2
\label{B.11}
\end{equation}
where $k_{nI}=\sqrt{\frac{2\mu}{\hbar^2}(E-\epsilon_{nI})}$.
The fusion cross section for an unpolarized target is then given by
\begin{equation}
\sigma_{fus}(E)=\frac{\pi}{(k_{n_iI_i})^2}\sum_{JML_i}
\frac{P^{J}_{L_iI_i}(E)}{2I_i+1}
=\frac{\pi}{(k_{n_iI_i})^2}\sum_{JL_i}
\frac{2J+1}{2I_i+1}
P^{J}_{L_iI_i}(E)
\label{B.12}
\end{equation}

We now introduce the path integral
representation of the penetration probability
for our coupled-channels problem.
Since heavy-ion fusion reactions are processes where two nuclei
approach close to each other,
we shall treat the radial component of the relative motion as
the macroscopic degree of freedom and its angular part and the
internal degrees of freedom as
environmental degrees of freedom.
The barrier transmission probability is then given by \cite{BT85}
\begin{eqnarray}
P^J_{L_iI_i}(E)&=&
\lim_{{R_i\rightarrow \infty}\atop{R_f\rightarrow -\infty}}
\biggl({P_i P_f \over {\mu^2}}\biggr)\int^\infty_0dT~e^{(i/\hbar)ET}
\int^\infty_0 d{\widetilde T}~e^{-(i/\hbar)E{\widetilde T}} \nonumber \\
&\times & \int {\cal D}\bigl[R(t)\bigr] \int {\cal D}\bigl[
{\widetilde R}({\tilde t}) \bigr] e^{(i/\hbar)\lbrack S_t (R,T)-S_t (
{\widetilde R},{\widetilde T}) \rbrack}\rho_M \bigl({\widetilde R}
({\tilde t}),{\widetilde T};R(t),T\bigr),
\label{10}
\end{eqnarray}
where
$P_i$ and $P_f$ are the classical momenta at the initial and the final
positions $R_i$ and $R_f$, respectively.
$S_t(R,T)$ is the action for the translational motion along a path
$R(t)$, and is given by
\begin{equation}
S_t(R,T)=\int^T_0 dt \left(\frac{1}{2}\mu\dot{R}(t)^2 -U(R(t))\right)
\end{equation}
The effects of the environmental degrees of freedom are included in the
two time influence functional $\rho_M$, which is defined by
\begin{equation}
\rho_M \bigl({\widetilde R}({\tilde t}),{\widetilde T};R(t),T\bigr)
=<(n_iL_iI_i)JM\vert \hat{u}^{\dagger}({\widetilde R}({\tilde t}),
{\widetilde T})\hat{u}(R(t),T)\vert (n_iL_iI_i)JM>
\end{equation}
with
\begin{equation}
i\hbar\frac{\partial}{\partial t}\hat{u}(R,t)=
\left[\frac{{\mbox{\boldmath $L$}}^2\hbar^2}{2\mu R^2}+H_0(\xi)
+V(R,\Omega,\xi)\right]\hat{u}(R,t).
\label{B.13}
\end{equation}
$\hat{u}(R,t)$ is the time evolution operator of the
environmental degrees of freedom
along a given path $R(t)$.
The formal solution of eq.(\ref{B.13}) can be written as
\begin{equation}
\hat{u}(R,t)=\hat{T}\exp\left[\int^t_0dt'\left(
\frac{{\mbox{\boldmath $L$}}^2\hbar^2}{2\mu R(t')^2}+H_0(\xi)
+V(R(t'),\Omega,\xi)\right)\right]
\label{17}
\end{equation}
where $\hat{T}$ is the time ordering operator.
Hereafter the time ordering is supposed to be properly treated
in all solutions of $\hat{u}$, and we shall not write it
explicitly.

We now assume that the angular part of the
translational motion is much slower than the radial
motion, and
replace the operator ${\mbox{\boldmath $L$}}^2$
in eq.(\ref{17}) by some c-number
$\Lambda(\Lambda+1)$\cite{HTBB95}.
This is a kind of sudden approximation and is exact
if there is no angular momentum transfer from the relative motion
between heavy-ions to nuclear intrinsic motion.
$\Lambda$ can be any c-number, though one often takes $\Lambda$ to be
the total angular momentum $J$.
If we denote the coordinate representation of $\Omega$ by
$\Omega'$\cite{HTBB95}, we get
\begin{eqnarray}
<\Omega'\vert \hat{u}(R,T)\vert (n_iL_iI_i)JM>&=&
\exp\left[\int^T_0dt'\left(
\frac{\Lambda(\Lambda+1)\hbar^2}{2\mu R(t')^2}+H_0(\xi)
+V(R(t'),\Omega',\xi)\right)\right] \nonumber \\
&&\times\sum_{m_L,m_I}<L_im_LI_im_I\vert JM>
Y_{L_im_L}(\Omega')
\vert\varphi_{n_iI_im_I}>
\label{B.17}
\end{eqnarray}

We next make a rotational coordinate transformation
in the whole space
to the coordinate system where
the $z$ axis is along the direction of the radial vector
${\mbox{\boldmath $R'$}}=(R',\Omega')=(R',\theta',\phi')$\cite{TAB92}.
We call the new coordinate system the rotating frame (RF) in order
to distinguish it from the space fixed frame(SF).
The operator for this coordinate transformation is given by
\begin{equation}
{\cal R}(\phi',\theta',0)=e^{i{\mbox{\boldmath $J$}}
\cdot {\mbox{\boldmath $\chi$}}(\Omega')/\hbar}
\end{equation}
In eq.(16) $ {\mbox{\boldmath $\chi$}}$ is the rotation vector which
specifies the direction and the magnitude of the rotation.
Note that the third Euler angle in this rotating frame
is zero.
Since the time evolution operator $\hat{u}(R,t)$
(see eq.(14)) does not change by rotation in
the no-Coriolis approximation,
we obtain
\begin{eqnarray}
<\Omega'\vert \hat{u}(R,T)\vert (n_iL_iI_i)JM>&=&
<\Omega'\vert {\cal R}^{-1}(\phi',\theta',0){\cal R}(\phi',\theta',0)
\hat{u}(R,T) \nonumber \\
&&\times {\cal R}^{-1}(\phi',\theta',0){\cal R}(\phi',\theta',0)
\vert (n_iL_iI_i)JM> \\
&=&\sum_K\bar{u}(R(t),T)<\Omega'=0\vert (n_iL_iI_i)JK>
D^J_{KM}(\phi',\theta',0) \\
&=&\sum_K\bar{u}(R(t),T)<L_i0I_iK\vert JK>\sqrt{\frac{2L_i+1}{4\pi}}
D^J_{KM}(\phi',\theta',0) \vert\varphi_{n_iI_iK}> \nonumber \\
\end{eqnarray}
where $D^J_{KM}$ is Wigner's $D$ function and
the time evolution operator in the rotating frame $\bar{u}(R,T)$
is defined as
\begin{eqnarray}
\bar{u}(R,T) &=&
\exp\left[\int^T_0dt'\left(
\frac{\Lambda(\Lambda+1)\hbar^2}{2\mu R(t')^2}+H_0(\xi)
+V(R(t'),\Omega=0,\xi)\right)\right] \\
&=&
\exp\left[\int^T_0dt'\left(
\frac{\Lambda(\Lambda+1)\hbar^2}{2\mu R(t')^2}+H_0(\xi)
+\sum_{\lambda>0}\sqrt{\frac{2\lambda+1}{4\pi}}
f_{\lambda}(R(t'))T_{\lambda0}(\xi)\right)\right]
\end{eqnarray}
In order to obtain eq.(19) we used
\begin{equation}
{\cal R}(\phi',\theta',0)\vert\Omega'>=\vert\Omega'=0>
\end{equation}
and
\begin{equation}
<\Omega'=0\vert Y_{L_im_L}>=\sqrt{\frac{2L_i+1}{4\pi}}
\delta_{m_L,0}
\end{equation}

The two time influence functional then becomes
\begin{eqnarray}
\rho_M \bigl({\widetilde R}({\tilde t}),{\widetilde T};R(t),T\bigr)
&=&\int \sin\theta'd\theta' d\phi'
<(n_iL_iI_i)JM\vert\hat{u}^{\dagger}({\widetilde R}
({\tilde t}),{\widetilde T})\vert
\Omega'>\nonumber \\
&&\times <\Omega'\vert
\hat{u}(R(t),T)\vert (n_iL_iI_i)JM> \\
&=&\sum_K\frac{2L_i+1}{2J+1}\vert<L_i0I_iK\vert JK>
\vert^2 \nonumber \\
&&\times <\varphi_{n_iI_iK}\vert \bar{u}^{\dagger}
({\widetilde R}({\tilde t}),
{\widetilde T})\bar{u}(R(t),T)\vert \varphi_{n_iI_iK}>
\label{B.25}
\end{eqnarray}
In obtaining eq.(25) from eq.(24) we used the orthogonality of the $D$
function
\begin{equation}
\int \sin\theta d\theta d\phi D^{J^*}_{K'M}(\phi,\theta,0)
D^{J}_{KM}(\phi,\theta,0)=\frac{4\pi}{2J+1}\delta_{K,K'}
\end{equation}
The time evolution operator in the rotating frame
$\bar{u}(R(t),t)$ obeys
\begin{equation}
i\hbar\frac{\partial}{\partial t}\bar{u}(R,t)=
\left[\frac{\Lambda(\Lambda+1)\hbar^2}{2\mu R^2}+H_0(\xi)
+\sum_{\lambda>0}\sqrt{\frac{2\lambda+1}{4\pi}}
f_{\lambda}(R)T_{\lambda0}(\xi)
\right]\bar{u}(R,t).
\label{B.27}
\end{equation}
Eq.(\ref{B.27}) shows that the $z$ component of the internal spin $m_I$
is conserved in the no-Coriolis approximation.
Since the wave functions with different values of $m_I$ never couple
to each other during the
reaction process, the dimension of the coupled-channels equations
is drastically reduced.
The effective Hamiltonian in eq.(\ref{B.27})
has the same form as that in the system where the internal
spin is zero.
The effects of the finite intrinsic spin enters only through
a scaling factor $\sqrt{\frac{2\lambda+1}{4\pi}}$ of the coupling strength.

 From eqs.(\ref{B.12}), (\ref{10}), and (\ref{B.25}),
the fusion cross section in the no-Coriolis approximation finally becomes
\begin{eqnarray}
\sigma_{fus}(E)&=&
\frac{\pi}{(k_{n_iI_i})^2}\sum_{JL_i}\sum_K
\frac{2L_i+1}{2I_i+1}
\vert<L_i0I_iK\vert JK>\vert^2 \nonumber \\
&&\times
\lim_{{R_i\rightarrow \infty}\atop{R_f\rightarrow -\infty}}
\biggl({P_i P_f \over {\mu^2}}\biggr)\int^\infty_0dT~e^{(i/\hbar)ET}
\int^\infty_0 d{\widetilde T}~e^{-(i/\hbar)E{\widetilde T}} \nonumber \\
&&\times  \int {\cal D}\bigl[R(t)\bigr] \int {\cal D}\bigl[
{\widetilde R}({\tilde t}) \bigr] e^{(i/\hbar)\lbrack S_t (R,T)-S_t (
{\widetilde R},{\widetilde T}) \rbrack}
<\varphi_{n_iI_iK}\vert \bar{u}^{\dagger}
({\widetilde R}({\tilde t}),
{\widetilde T})\bar{u}(R(t),T)\vert \varphi_{n_iI_iK}> \nonumber \\
&&
\label{30}
\end{eqnarray}
If the initial value of the internal spin is zero,
the initial angular momentum for the relative motion $L_i$
equals $J$, and
the summation in eq.(\ref{30}) becomes simple.
The fusion cross section in that case can be calculated by treating as
though the relative motion couples to a spinless mode
of excitation except for
the scaling factor $\sqrt{\frac{2\lambda+1}{4\pi}}$ mentioned above.
If the initial spin of the internal motion is finite,
the influence functional is obtained by first calculating it for a fixed
$K$-quantum number, and then by taking sum with the weight following
Clebsch-Gordan coefficients.

Before closing the paper, we wish to comment on
the applicability of the no-Coriolis approximation.
It is known that the no-Coriolis approximation
cannot be applied when a long range force, such as the
Coulomb interaction, is involved[16,19--21].
%\cite{ALS93,TMBR91,AA94,GCAN94}.
Heavy-ion fusion reactions are governed by
the behavior of the wave functions in the small region
near the Coulomb barrier.
The no-Coriolis approximation is, therefore , considered
to be a good approximation.
On the other hand, if one is interested in the
angular distributions of elastic and inelastic scattering,
the no-Coriolis approximation fails to give the correct
scattering phase shifts.
In order to cure the problems in such cases,
a prescription of renormalizing the
coupling strength has been proposed by several authors
\cite{TMBR91,AA94,GCAN94}.
The way of renormalization is, however, not unique and
this problem is still unsettled.

In summary, we used the path integral method to reformulate
the coupled-channels problems in the no-Coriolis approximation.
We first ignored the change of the centrifugal potential due to
an intrinsic excitation.
We then introduced a rotational coordinate transformation
into the coordinate system where
the $z$ axis is along the direction of the radial vector of
the relative motion.
We have thus shown that the fusion cross section can be
calculated by treating nuclear intrinsic motions as though they
do not carry a finite angular momentum. The finite multipolarity
of nuclear intrinsic excitation appears
as a scaling factor in the coupling strength.
Though these results have already been obtained by different methods,
the advantages of our path integral formulation are that
we can easily apply the same technique to cases where the initial internal
spin is not zero, and also where the internal angular momentum is not the
spin but the orbital angular momentum.
The effect of transfer reactions on
heavy ion fusion reactions is one such problem \cite{EL89}.

\medskip

The work of K.H. was supported by Research Fellowships
of the Japan Society for the Promotion of Science for
Young Scientists.
That of J.R.B. was supported by a Japan Society for Promotion
of Science Postdoctoral Fellowship for Foreign Researchers in Japan.
A.B.B. acknowledges a fellowship from the Japanese Society for Promotion
of Science.
This work was supported in part by the Grant-in-Aid for General
Scientific Research,
Contract No.06640368, in part the Grant-in-Aid for Scientific
Research on Priority
Areas,Contract No.05243101,from the Japanese Ministry of Education,
Science and Culture, in part by the U.S. National Science
Foundation Grant No. 9314131 and 9303041 and in part by the
U.S. Department of Energy Grant DE-FG05-94ER40827.

\end{document}